\documentclass[twoside,twocolumn]{article}

\usepackage{titlesec}

\usepackage{blindtext} % Package to generate dummy text throughout this template 

\usepackage[sc]{mathpazo} % Use the Palatino font
\usepackage[T1]{fontenc} % Use 8-bit encoding that has 256 glyphs
\usepackage{microtype} % Slightly tweak font spacing for aesthetics

\usepackage[english]{babel} % Language hyphenation and typographical rules

\usepackage[hmarginratio=1:1,top=32mm,columnsep=20pt]{geometry} % Document margins
\usepackage[hang, small,labelfont=bf,up,textfont=it,up]{caption} % Custom captions under/above floats in tables or figures
\usepackage{booktabs} % Horizontal rules in tables

\usepackage{lettrine} % The lettrine is the first enlarged letter at the beginning of the text

\usepackage{enumitem} % Customized lists
\setlist[itemize]{noitemsep} % Make itemize lists more compact

\usepackage{abstract} % Allows abstract customization
\usepackage{titlesec} % Allows customization of titles
\renewcommand\thesection{\Roman{section}} % Roman numerals for the sections
\renewcommand\thesubsection{\roman{subsection}} % roman numerals for subsections
\titleformat{\section}[block]{\large\scshape\centering}{\thesection.}{1em}{} % Change the look of the section titles
\titleformat{\subsection}[block]{\large}{\thesubsection.}{1em}{} % Change the look of the section titles

\usepackage{fancyhdr} % Headers and footers
\usepackage{titling} % Customizing the title section

\usepackage{hyperref} % For hyperlinks in the PDF

\pretitle{\begin{center}\Huge\bfseries} % Article title formatting
\posttitle{\end{center}} % Article title closing formatting
\title{Machine Learning for Mars Exploration} % Article title
\author{%
\textsc{Ali Momennasab}% Your name
}
\date{} % Leave empty to omit a date
\renewcommand{\maketitlehookd}{%
\begin{abstract}
\noindent {Risk to human astronauts and interplanetary distance causing slow and limited communication drives scientists to pursue an autonomous approach to exploring distant planets, such as Mars. A portion of exploration of Mars has been conducted through the autonomous collection and analysis of Martian data by spacecraft such as the Mars rovers and the Mars Express Orbiter. The autonomy used on these Mars exploration spacecraft and on Earth to analyze data collected by these vehicles mainly consist of machine learning, a field of artificial intelligence where algorithms collect data and self-improve with the data. Additional applications of machine learning techniques for Mars exploration have potential to resolve communication limitations and human risks of interplanetary exploration. In addition, analyzing Mars data with machine learning has the potential to provide a greater understanding of Mars in numerous domains such as its climate, atmosphere, and potential future habitation.  To explore further utilizations of machine learning techniques for Mars exploration, this paper will first  summarize the general features and phenomena of Mars to provide a general overview of the planet, elaborate upon uncertainties of Mars that would be beneficial to explore and understand, summarize every current or previous usage of machine learning techniques in the exploration of Mars, explore implementations of machine learning that will be utilized in future Mars exploration missions, and explore machine learning techniques used in Earthly domains to provide solutions to the previously described uncertainties of Mars. } % Dummy abstract text - replace \blindtext with your abstract text
\end{abstract}
}

\begin{document}

% Print the title
\maketitle

%----------------------------------------------------------------------------------------
%	ARTICLE CONTENTS
%----------------------------------------------------------------------------------------

\section{Introduction}

\lettrine[nindent=0em,lines=3]{M} achine learning is a field of artificial intelligence that studies algorithms that learn, or improve their performance, using collected data.
\noindent{The ability of machine learning algorithms to capture, analyze, and describe complex data patterns makes them a valuable tool for studying various phenomena across numerous industrial and scientific domains [1]. Machine learning techniques have also been applied successfully to spacecraft and towards analyzing space-collected data, as detailed in section 3. 
} % Dummy text

\noindent{Space missions are often carried out remotely because of the risk to human astronauts, the costs of maintaining astronaut safety and health, and distance from Earth. Some locations in space are inaccessible to human exploration due to harsh, dangerous, remote, and inhospitable conditions. In addition, large distances result in two obstacles: extremely long communication times between spacecraft and Earth lead to wasted time and productivity while messages transmit between spacecraft and Earth, and a limit on data transmitted due to bandwidth limitations at large distances leads to discarding data in order to adhere to the low bandwidth. Due to these risks and drawbacks, autonomous robotic agents such as rovers are advantageous for space exploration [2]. Additionally, Mars exploration is a data-rich field, with future missions set to collect larger and more detailed datasets than before, significantly increasing the total data available and the rate of new observations. These data analysis challenges can be addressed by machine learning data analysis techniques.} % Dummy text

%------------------------------------------------

\section{Features of Mars beneficial to explore and understand}
This section will summarize basic atmospheric and geological features of Mars, which will be referenced in future sections, to provide a basic understanding of Mars. The introduction will define general features or phenomena of Mars, and obstacles that hinder human exploration. Uncertainties, or features or phenomena of Mars that are undiscovered or unknown, will be described to establish why they would be beneficial to resolve.

\subsection{Atmospheric Features}

\subsubsection{Overview of the atmosphere of Mars}

The atmosphere of Mars at the surface is mainly composed of carbon dioxide (95\%), molecular nitrogen (2.6\%), argon (1.9\%), molecular oxygen (0.16\%), and carbon monoxide (0.06\%), additionally containing differing trace amounts of methane. These gases have been observed by the Sample Analysis at Mars (SAM) instrument onboard the Curiosity Rover [5], [9] to fluctuate in quantity from their seasonal averages depending on the season, with molecular oxygen varying by 13\% on average [9].

\noindent{The Martian atmosphere is an oxidizing atmosphere. O2 within the atmosphere corrodes the iron oxide of the Martian surface material, resulting in the signature red color of the planet. The Martian atmosphere similarly corrodes certain metals and degrades certain materials. For example, the wheels of the Curiosity Rover deteriorated after spending time on Mars partially because of the oxidizing atmosphere [5], [10].}

\noindent{The Mars atmosphere, which is much thinner than the atmosphere of Earth, offers little protection against the radiation-emitting galactic cosmic rays and solar energy particles exuded by the sun [5], [6] and prevents heat retention [5]. These factors, along with a massive distance away from the sun, result in a frigid Martian surface with an average surface temperature of ~220K [5], [7]. In addition to low temperatures, the Martian exterior frequently experiences dust storms and surface winds [5]. These harsh atmospheric conditions of the Martian surface altogether make current human exploration hazardous.}

\noindent{Clouds are another weather feature of Mars, which have been observed on Mars by the light detection and ranging instrument on the Phoenix mission [11], by the Mars Global Surveyor Mars Orbiter Camera [7], and by the Opportunity rover [12]. Martian clouds are similar to cirrus clouds of Earth [7], [11]. However, the clouds observed by the Phoenix mission and the Mars Orbiter Camera were water ice clouds, while the clouds captured by the Opportunity rover were likely dry ice clouds. Additionally, the clouds captured by the Opportunity rover were at a higher altitude than typical Martian clouds [12]. Colder temperatures occur at higher altitudes, thus increasing the likelihood that the Opportunity-observed clouds were frozen carbon dioxide.}

\noindent{Mars has an atmospheric pressure of ~610 Pa [13]. For reference, Earth has an atmospheric pressure of 101352.9 Pa, so the atmospheric pressure of Mars is less than 1\% of the atmospheric pressure of Earth. Due to low atmospheric pressure and low temperatures, liquid water cannot currently exist at the surface of Mars [5], [14]. The lack of liquid water on the surface of Mars contributes to the risk and difficulty of human exploration of Mars.}

\subsubsection{Fluctuating Amounts of Oxygen in the Martian Atmosphere}

The Tunable Laser Spectrometer of the Sample Analysis at Mars (SAM) instrument onboard Curiosity has detected that the amount of oxygen on Mars rises by 30\% more than predicted seasonal patterns estimated during Martian springs and summers for unknown reasons [9]. Understanding why the oxygen of Mars fluctuates would contribute towards understanding how the Martian surface and atmosphere interact and might indicate habitability or presence of life [9].

\subsubsection{Unknown source of methane in the Martian atmosphere}
An unknown process produces the methane in the Martian atmosphere measured by SAM’s Tunable Laser Spectrometer, possibly through microbial life or abiotic processes [15], [17] including serpentinization, the process where olivine or pyroxene is heated under pressure and reacts with water and carbon dioxide to create methane and the mineral serpentine [17]. Methane produced by microbial processes would indicate the presence of microbial life on Mars. Methane produced by serpentinization would signify additional evidence of subsurface liquid water, further substantiating the previous evidence of subsurface water detected under the southern polar cap by scans from the Mars Radar for Subsurface and Ionospheric Sounding onboard the Mars Express spacecraft [18]. Thus, understanding the source of and patterning of methane in the Martian atmosphere would contribute to the search for signs of habitability on Mars, improve our general understanding of Mars, and further assure the presence of subsurface water on Mars.

\subsubsection{Martian cloud distribution}
Martian clouds typically form during the coldest part of the Martian year, where the planet is at the furthest distance from the sun in its elliptical orbit. However, cameras onboard the Opportunity rover have recently captured Martian clouds earlier in the Martian year than predicted and at higher altitudes than normal [14]. As the positioning and timing of Martian clouds is now different from predictions, understanding the spatial distribution and timing of Martian clouds would result in a greater understanding of the Martian hydrological cycle and how the Martian atmosphere operates [12], [14]. 

\subsubsection{Mars weather analysis}
Mars has a diverse weather system, with general features including but not limited to temperature, humidity, dust storms, surface winds, and pressure [4], [5], [7], [49]. Analyzing these general features of Martian weather would allow for predictive models of the future and past Martian climate. These models would increase knowledge of the Martian climate and atmosphere, strengthen or weaken arguments regarding the future habitation and colonization of Mars, and contribute to the overall search for future human habitability on Mars [49].

\subsection{Geological Features}
\subsubsection{Overview of the geological features of Mars}
The uppermost layer of Mars is covered by regolith composed of red soil, varying amounts of fractured rock, and dust [19]. The regolith can be abrasive if it is rocky, which is a partial reason for wheel degradation of the Curiosity Rover [10]. Martian regolith covers craters, canyons, extinct volcanoes, mountains, dry lakebeds, dry channels, and dry valley networks, as well as smaller landforms such as dunes and gullies. A majority of the craters that span across Mars are impact craters formed by asteroid or comet crashes. In addition, common to the terrain of Mars are valley networks, which are systems of dry channels that traverse the Martian surface. They indicate the presence of previously active but now dry water networks [20]. As previously mentioned, liquid water cannot exist on the surface of Mars due to the low atmospheric pressure of the planet. However, two permanent polar ice caps consisting of water ice and carbon dioxide are located at opposing poles of Mars. Within both polar ice caps are polar-layered deposits built from layers of water ice and dust, captured by Mariner 9 and Viking missions [21]. There are also smaller ice patches on the surface of Mars, such as Korolev Crater.  

\subsubsection{Martian valley networks and their stream orders}
The valley networks traversing the Martian surface remain largely unmapped due to a lack of spatial integration and organization in typical drainage mapping algorithms [22]. A few valley networks have been mapped manually, which is time-consuming and labor-intensive, but most valley networks remain uncharted. Further mapping of valley networks would improve understanding of the history of liquids on Mars and the climate of Mars [20].

The term stream order represents a positive integer that conveys the complexity of channel drainage, indicating the method and duration of water provided. There are numerous stream-ordering methods, but the most commonly used is the method of Strahler [22]. The stream ordering method of Strahler designates an order of k=1 to streams that have no tributaries (a tributary is a stream or river that flows into the main stem river, a larger stream, or a lake) or to the outermost streams of a stream network. When streams of the same order meet, they combine to form a stream with an order of k+1. However, when streams of different orders meet, the resultant stream order will be the largest stream order value of the converging streams [22]. Calculating the stream order of Martian valley networks would contribute towards a greater understanding of the presence of liquids in the past and hence greater comprehension of the climate of Mars.

\subsubsection{Discovering impact craters on the surface of Mars}
Impact craters can form at any time due to an asteroid or comet collision with Mars. For example, the HiRise camera of the Mars Reconnaissance Orbiter recently captured new impact craters in 2019 and 2021. It would be beneficial to discover and map all existing and newly formed impact craters because analysis of the spatial statistics of impact craters results in information about the geological properties and processes of the Martian surface [23]. However, it is difficult to manually analyze Mars surface imagery to map impact craters because it is time-consuming and labor-intensive to analyze thousands of Mars surface imagery to discover impact craters, which are prone to forming randomly. 

\subsubsection{Cyclicity of layered terrain at Mars polar caps}
The layer depth, patterning, and frequency of the polar-layered deposits at the Martian polar caps appear to be stochastic from images captured by the Mariner 9 and Viking missions. This leads to the question of whether the disposition of the polar-layered deposits is cyclic [21]. Information about the Martian climate could be revealed from the polar-layered deposits if any cyclicity or pattern is discovered in the formation of layered terrains at the polar caps [26].

Unconformities develop when the Martian polar cap layers interrupt or overlay other layers due to layer erosion [27]. Images captured by the Viking mission display several unconformities spanning the PLDs at the Martian polar caps [21]. The erosion of PLDs that causes unconformities indicates previous climate change [21], so mapping and patterning the polar unconformities could provide information about the recent Martian climate and its evolution, along with how the PLDs changed with time [24], [26].

\section{Machine learning previously or currently used for Mars exploration or data analysis}
This section will summarize applications of machine learning techniques utilized on either Mars (in-situ) or Earth (ex-situ) mainly to prioritize the collection of vital samples, analyze terrain and locate target terrain features, and analyze data collected on Mars. 

\subsection{In-situ}
\subsubsection{OASIS and AEGIS for rock analysis and terrain features detection}
The Onboard Autonomous Science Investigation System for Opportunistic Rover Science (OASIS) framework operates onboard Mars rovers to identify, analyze, and prioritize data for downlinking to Earth. OASIS has three main components: extracting features from images, analyzing and prioritizing data, and planning and scheduling new sequences based on the data analysis [8], [26]. Terrain features are identified in images using machine learning algorithms [8], [26]. Feature extraction from images occurs after terrain target identification, where the physical properties of the terrain are classified using factors such as albedo (a measure of the reflection of a surface), size, and shape variation. After terrain target identification and classification, prioritization of data acquisition occurs when humans manually configure targeting algorithms to assign values of importance to specific terrain features. OASIS then compares the designated most important terrain features to the terrain features present in the terrain imagery and prioritizes the terrain features that most closely match the target features by examining the distance of the extracted feature vectors from the weighted feature vector.

The Autonomous Exploration for Gathering Increased Science system (AEGIS) is a component of the OASIS autonomous framework that provides automatic targeting for remote sensing instruments on Mars rovers and data analysis of images collected for the identification and targeting of features of interest [2], [8]. AEGIS uses the terrain feature identification of OASIS to guide the ChemCam laser of the rover, automatically selecting then vaporizing target terrain features to analyze emanated plasma. The autonomous control of the ChemCam increases the pace of data acquisition and lessens lost time because AEGIS removes the need for human commandeering [8], [27]. AEGIS is also present on the Perseverance Rover, controlling the SuperCam of the rover to identify the chemical composition of rocks and soils [27].

\subsubsection{ENav and ACE for rover path plannning}
Enhanced AutoNav, known as ENav, is the navigation software used by the Perseverance rover. ENav designates and ranks a list of paths for the Perseverance rover, then uses the Approximate Clearance Evaluation, or ACE, algorithm to validate the safety of the highest-ranked routes [28], [29]. ACE approximates the lowest and highest possible height of each rover wheel with measurements from rover sensors of the suspension angle of each rover wheel and the terrain, then predicts the safety of a path using stimulated worst-case scenarios of wheel positioning [29]. Because of automated navigation, the time spent for signals from Earth to reach the rover during manual operation is eliminated, allowing for extensive coverage of the Mars surface and reduced lost time. However, ACE will evaluate the entirety of the list to find a feasible path if it cannot determine a suitable path from the array of best paths, which is time-consuming and computationally demanding [29]. Reference [28] presents a machine learning model tested using Monte Carlo trials to have less computation time per amount of ACE evaluations, increase path efficiency, and preserve or improve the rate of successful traverses without compromising the detection of unsuitable terrain of ACE.	

\subsubsection{OASIS dust devil and cloud detection in Mars rover-captured surface imagery}
The OASIS software (also present in AEGIS, see section 2.1.1) onboard Mars Exploration Rovers (MER) downlinks portions of images or full images only containing targeted or desired features after analyzing rover Navcam imagery of dust devils and clouds [30]. Before implementing OASIS onboard MER rovers, the rovers captured sets of images at fixed times, and then downloaded the entirety of the image sets. There was not much opportunity for data acquisition of targeted, and usually rare, events such as dust devils or clouds because of image capturing at fixed times. The implementation of this command, known as WATCH, allowed for a larger period for capturing targeted phenomena and a reduction in the amount of relayed images, hence resulting in immense bandwidth savings [30].

\subsection{Ex-situ}
\subsubsection{COSMIC discovering of Mars impact craters}
The Capturing Onboard Summarization to Monitor Image Change (COSMIC) project is currently in development to automatically analyze data onboard Mars spacecraft and notify scientists when anything noteworthy occurs or changes in order to eliminate bandwidth limitations at large distances [31]. As a part of COSMIC, JPL scientists created an automatic impact crater classifier, which analyzes images captured by the Martian Reconnaissance Orbiter (MRO) to discover impact craters on the surface of Mars. As described in section 1.2.3, new impact craters can form randomly. Before utilizing machine learning techniques for crater discovery, crater discovery initially resulted from time-consuming and labor-intensive manual human analysis of MRO imagery of the Martian surface. The utilization of machine learning methods for analyzing MRO-captured imagery to discover impact craters resulted in easier discovery of smaller impact craters, less wasted time manually analyzing images, and an increase in crater discoveries [31].

\subsubsection{CRISM Mars crater mineral discovery}
The Compact Reconnaissance Imaging Spectrometer for Mars (CRISM), located on the Mars Reconnaissance Orbiter, uses detectors to search for aqueous mineral residue. Previous CRISM discovery of aqueous minerals on the Martian surface was monumental towards improved understanding of Mars [32]. However, mainly commonly occurring minerals and mineral phases, not secondary or accessory phases, were discovered by CRISM [32]. Reference [32] have developed machine learning methods for automatic mineral discovery of less common minerals in CRISM-acquired data, resulting in mineral discoveries that suggest the existence of water in the Jezero crater landing site and the Northeast Syrtis region.

\subsubsection{SPOC and AI4Mars}
The Soil Property and Object Classification (SPOC) software uses machine learning to identify terrain types and features in orbital and ground-based images [33]. SPOC is a component of the Machine learning-based Analytics for Automated Rover Systems (MAARS) effort, a system of autonomous algorithms designed to improve the safety and productivity of future rover missions (section 2.2.3 summarizes the other components of MAARS) [34]. SPOC analysis of images captured by the Navcam of the Curiosity rover to identify areas of slippage on Mars terrain and images captured by the HiRISE camera to determine the traversability of potential landing sites for the Mars 2020 Rover mission. SPOC has been successful for these purposes and will become more successful after refinement using the AI4MARS dataset [35]. AI4Mars is a large dataset totaling ~35K high-resolution images taken on the surface of Mars from the Opportunity, Spirit, and Curiosity rovers, with approximately ten people labeling each image in the dataset to ensure that each image is high quality [35]. Training SPOC using the AI4MARS dataset increases the accuracy of SPOC, and future improvements to the algorithm will result in further improved traversability analysis and terrain-feature discovery [33].

\subsubsection{Mars express power-predicting algorithm}
The Mars Express (MEX) spacecraft of the European Space Agency has been exploring the surface of Mars and performing science operations from orbit since 2004 [36], [37], such as finding evidence of the presence of subsurface Martian water using its Mars Radar for Subsurface and Ionospheric Sounding (described in 2.1.3) [18]. The MEX orbiter is powered by electricity generated by its solar panels, stored in batteries to be used when the sun is not present [36], [37]. The autonomous thermal system of MEX, which maintains the temperature of every instrument onboard MEX, uses a majority of the total generated electric power, leaving only a portion of energy for scientific operations [36], [37]. In order to effectively distribute the total energy between the thermal system and planned scientific procedures, the power consumption of the thermal system, which varies depending on factors such as spacecraft heat and instrument heat, must be predicted to allocate the remaining energy towards scientific operations. Reference [37] has presented a machine learning model trained with three Martian years of telemetry data and thermal system data, including the measured electric current of the thermal system, to predict the values of the electric current used by the thermal system. However, the raw data cannot initially train machine learning algorithms because it consists of data in formats that machine learning algorithms cannot process (ex. text) and has incompatible time resolutions. The raw data is transformed and then employed by multi-target regression to produce accurate and efficient models of predicted thermal energy consumption using a common set of inputs. This machine learning technique can also be utilized onboard Mars rovers to predict rover energy consumption, similar to the proposed MER rover driving energy prediction VeeGer algorithms described in section 4.1.1.

\section{Novel, theoretical, and future applications of machine learning for Mars exploration}
This section will summarize applications of machine learning techniques in current development for future utilization on Mars (in-situ) or Earth (ex-situ). In addition, this section will explore machine learning techniques applications of Earthly domains with amenability to Mars exploration and data analysis, as well as novel machine learning applications for Mars exploration and data analysis. 

\subsection{In-situ}

\subsubsection{MAARS framework components: SCOTI, RAND, VEEGER, CAAPS, OBKS, AND P-ACE}
SPOC, a component of the autonomous MAARS framework described in section 2.2.3., has been utilized on the Curiosity Rover [35]. However, the remaining majority of the MAARS algorithms, including SCOTI, RAND, VeeGer, p-ACE, OBKS, and CAAPS, have not been utilized onboard MER rovers currently but will be used on future MER missions.

\paragraph{SCOTI}
The Scientific Captioning Of Terrain Images (SCOTI) model automatically creates captions for pictures of the Martian surface [34], [38]. Reference [34] describes in detail how SCOTI is trained. Any imagery of the surface of Mars is input into SCOTI, and SCOTI outputs an English caption that describes the geographic features of the image [34], [38]. SCOTI can prioritize the downlink of terrain imagery with captions containing certain words corresponding to desired terrain features onboard MER rovers by uplinking words and phrases containing terrain features of interest to the rover. Additionally, scientists can use downlinked SCOTI-created captions to select images containing specific geologic features of interest to be downlinked by MER rovers and summarize all features present in terrain imagery. Although SCOTI utilization is mainly for onboard MER rovers, SCOTI can also be used on Earth to filter millions of images of the Martian surface for specific features based on the captions of the images through a text-based query. SCOTI reduces the volume of downlinked images to keep in limit with the downlink bandwidth of an MER rover, allows for the downlink of only images containing features of interest, and saves scientists from filtering through millions of terrain images in search of features of interest [34], [38].

\paragraph{RAND}
The Resource-Aware planner for Non-stop Driving (RAND) algorithm, in current development, will provide onboard rover path planning for as little driving time as possible. RAND evaluates Monte Carlo simulations of paths towards a designated goal target, randomly sampling factors such as the location of the rover, time of day, and the amount of rover energy during its calculations [34]. The algorithm ranks the numerous trajectory simulations, and the best paths are compressed and uplinked to the rover. The network of remembered trajectories is then decompressed on the rover and used by a Mars rover to navigate the surface of Mars whenever the rover must alter its current path or has a significant change in energy level. For an example of how effective RAND compression is, [34] demonstrated that RAND compressed a search space of the Jezero crater landing site to only three trajectories, compressing the original ~1 gigabyte of raw planning data to only <1 gigabyte of data. RAND is also unique in the sense that more data leads to less computational strain. If there are more networks uploaded to the rover, there will be more trajectory options for the rover to select, and the rover will have less computational strain trying to choose a feasible path. However, the rover will have a large computational load when selecting the most suitable path option from a small array of networks, as lesser options might mean fewer good paths and more computing to determine the best route. RAND can eliminate reliance on Earth planning for rover navigation, save space, and conserve rover energy by choosing the most efficient path towards a target location, increasing driving time [34]. 

\paragraph{VeeGer}
The proposed Vision-based Driving Energy Prediction algorithm will predict the driving energy of future MER rovers before traveling and potential slippage of the Martian surface with the image identification of SPOC (see 2.2.3) and obstacle detection with sensors on rovers such as stereo vision [34], [39]. Because future MER rovers are likely to be powered using solar panels, a predictive energy algorithm would be vital for rover path planning when the sun is not present. There are two VeeGer machine learning approaches, VeeGer-EnergyNet and Veeger-TerramechanicsNet. Veeger-TerramechanicsNet predicts the terramechanics parameters of the rover using a convolutional neural net that predicts wheel-terrain interactions using depth images and RGB and then calculates energy consumption based on the simplified terramechanics model [39]. VeeGer-EnergyNet uses only RGB and depth images captured by rover cameras to estimate energy consumption, skipping the terramechanics estimations of the VeeGer-TerramechanicsNet model [39]. However, both of these machine learning models have in common that they both rely on images to predict energy consumption. Testing both the algorithms on the Athena test rover of JPL determined that the VeeGer-TerramechanicsNet model predicted energy consumption with greater accuracy than the VeeGer-EnergyNet model [34]. VeeGer is likely to be applied to future MER rovers under the MAARS autonomous system described in 2.2.3, combined with future solar-energy generation calculation algorithms and path-planning algorithms such as ENav (see 2.1.2) for energy-optimal autonomous driving. 

\paragraph{CAAPS, OBKS, and p-ACE}
The MAARS autonomous framework provides three obstacle-checking algorithms: Approximate Clearance Evaluation (ACE, see 2.1.2), Optimization-Based Kinematic Settling (OBKS), and p-ACE, a probabilistic extension of ACE. MAARS uses the Context-Aware Adaptive Policy Selection (CAAPS) algorithm to select the optimal planning algorithm for a current environment that would result in the lowest path generation time without compromising safety [34]. 

As mentioned in 2.1.2, ACE determines the feasibility of an area for Mars rover traversal using the worst-case scenarios of the suspension system of a rover and the height of the terrain. However, the intrinsic conservative approach of ACE often results in ACE designating feasible areas as infeasible [34], [40]. A proposed extension of the ACE algorithm, p-ACE, uses real-time probability distributions in real-time rather than the restrictive and conservative bounds of ACE to evaluate the safety of terrain without compromising rover safety. Unlike the deterministic validation of ACE that estimates terrain safety using worst-case scenarios despite their unlikelihood of occurring, the probabilistic validation of p-ACE determines the probability of a worst-case scenario resulting in a collision before assessing terrain feasibility [34], [40]. 

The OBKS planner minimizes contact between rover wheels and terrain with less conservatism than ACE, similar to p-ACE. OBKS solves a local optimization problem modeled as a least-squares problem subject to pose constraints on joint angles as determined by rover design limits to minimize contact between terrain and rover wheels [34]. Because OBKS calculations expect to result in a close representation of the exact rover position for a given location on a heightmap, the interval of uncertainty is smaller, making the path planner more conservative than ACE. Compared to ACE, OBKS has a greater success rate and lower path selection time in areas with more terrain complexities, such as increased rock abundance, and has increased efficiency due to less conservatism, but has a larger time per query [34]. 

The CAAPS algorithm selects the most optimal planning algorithm between OBKS, ACE, or p-ACE depending on the environment and state of a Mars rover [34]. Depending on the environmental factors of the rover, CAAPS selects a planner that minimizes the energy needed to compute the planner and path generation time because different levels of environmental factors lead to varying path generation time between the three planners. For example, ACE path generation is the slowest and OBKS the fastest when rock abundance is high, and vice versa when rock abundance is low [34]. In such a scenario, the path selection of CAAPS, which combines the planners to select the most efficient path generation given the environmental factors, reduces path generation time with greater efficiency than a single planner.

\subsubsection{Mars oxygen detection and analysis}
As described in section 2.1.2, The Sample Analysis at Mars (SAM) instrument onboard the Curiosity rover measures the chemical composition of major atmospheric species, including oxygen, using the Tunable Laser Spectrometer. SAM has observed amounts of oxygen that fluctuate significantly during Martian summers and springs for unknown reasons. Understanding the cause of the oxygen fluctuation would reveal information on how the Martian surface and atmosphere interact and contribute to the search for habitability or life [9].

Machine learning techniques are amenable to detecting and patterning oxygen, albeit not utilized on Mars yet. Reference [41] has developed an extreme machine learning model (ELM) that determines the robustness of oxygen using data collected by a tunable laser spectrometer, which is the same technology that SAM uses to observe oxygen in the Martian atmosphere. If applied to the oxygen data of SAM, the ELM model of [41] would reduce data uncertainty and allow for greater accuracy of oxygen data that would result in a greater understanding of how the Martian surface and atmosphere interact and the habitability of Mars. In addition, machine learning techniques of pattern recognition are also amenable to SAM oxygen data. Recognizing a pattern in the abundance of oxygen of Mars that would indicate the presence of a biological or geological process can exhibit interactions between the Martian surface and atmosphere, and ascertain the habitability of Mars [9].

\subsubsection{Mars methane detection and analysis}
As described in 2.1.3, the source of methane measured by the Tunable Laser Spectrometer of Curiosity is currently unknown, but theories range from abiotic processes like serpentinization [15], [16] to microbial life [15]. Serpentinization would provide further evidence towards subsurface liquid water, and microbial life would indicate life on Mars. Thus, understanding how Martian methane produces can provide immense information about the planet and its habitability.

Machine learning techniques can identify the source and patterning of the Martian Methane. Reference [42] has developed a probabilistic machine learning algorithm that predicts high-emitting sites of methane with 70\% accuracy [42]. The algorithm could predict the source of methane on Mars if utilized with methane data from the Tunable Laser Spectrometer. With the origin of the methane, scientists can determine the process used to create Martian methane, which could potentially indicate the existence of life on Mars or provide additional evidence for the presence of subsurface liquid water.

\subsubsection{Automatic onboard pixel classification of Mars rover-captured surface imagery}
As described in section 3.1.1, OASIS is currently onboard MER rovers to analyze photographic data captured by the MER’s Navcam and select images with features of interest to be downlinked back to Earth, and help with the rover’s navigation. OASIS is effective, but a downside is that downlinking high-quality images useful for scientists to analyze requires significant rover energy [43]. Reference [43] proposed an extreme machine learning (ELM) pixel classifier algorithm to reduce the amount of rover energy needed to downlink high-quality images to Earth, as well as increase the accuracy and quality of downlinked images. The proposed ELM algorithm improved pixel identification accuracy and run time and reduced rover energy needed to downlink images [43], allowing for a potential upgrade of OASIS without compromising functionality.

\subsubsection{Martian cloud patterning}
As described in 2.1.4, Martian clouds as of recent, captured by the Opportunity rover’s Navcams, have deviated from typical formations and have begun to form earlier during the Martian year and at irregularly high altitudes [12]. Recognizing the spatial distribution and timing of Martian clouds could result in a greater understanding of how the Martian atmosphere and hydrological cycle operate [12].

Machine learning techniques can identify patterns of the spatial distribution and timing of Martian clouds. For example, [43] has demonstrated the effectiveness of using machine learning to build models of clouds based on Earth cloud imagery. Although [43] uses machine learning to model Earth clouds, the same cloud modeling algorithms can be amenable to image data of Martian clouds captured by cameras on MER rovers. If the distribution and timing of Martian clouds were to be modeled by machine learning algorithms, a greater understanding of the Martian atmosphere, as well as the Martian hydrological cycle, can be attained [12].

\subsection{Ex-situ}
\subsubsection{Mars weather analysis}
Analyzing numerous features of Martian weather, including temperature, humidity, dust storms, surface winds, and pressure [4], could increase knowledge of the Martian climate and atmosphere, provide evidence towards debates regarding the future habitation and colonization of Mars, and support the search for habitability, as summarized in section 2.1.5. Machine learning techniques are amenable to analyzing the numerous factors that compose the Martian weather, as demonstrated by [49], who was the first to explore the analysis of Martian weather using several machine learning models. Using Mars weather data, [49] tested machine learning models including convolution neural networks, gated recurrent units, long short term memory (LSTM), stacked LSTM, and CNN-LSTM to determine that LSTM provided the best analysis with the least error [49]. With significantly more data available to train the model and the usage of more models, future analysis of Martian weather would result in a greater exploration of Martian weather and the identification of conditions that would contribute to the search for sustainability and habitation on Mars [49].

\subsubsection{Mars valley network mapping and stream ordering}
As detailed in 2.2.2, numerous valley networks that indicate the past presence of running water on the Martian surface cross the surface of Mars. However, because current network mapping algorithms lack spatial integration and organization and it is labor-intensive to map each network manually, most Martian valley networks remain unmapped [20]. Further valley network mapping would increase understanding of the history of liquids on Mars and hence the climate of Mars. Machine learning algorithms can effectively map Martian valley networks, as demonstrated by [20]. Reference [20] proposes a machine learning algorithm that maps Martian valley networks using images captured from the Mars Odyssey Spacecraft’s THEMIS camera. The valley networks mapped by machine learning were of better quality than manually mapped valley networks, resulting in more accurate valley network mapping and less wasted time. Therefore, using machine learning to map Martian valley networks can increase knowledge of the climate and history of the past surface liquid of Mars while creating higher-quality maps than manually created mapped networks [20].

The calculation of the stream order of each Martian valley network would also contribute towards a greater understanding of the presence of past surface liquids, and hence greater comprehension of the climate of Mars. Machine learning algorithms can calculate the stream order of valley networks, as demonstrated by [22]. Reference [22] proposes a machine learning framework that calculates the stream order of a valley network using the stream ordering method of Strahler that is more efficient and less time-consuming than previous algorithms. Martian valley network data can train the stream-ordering framework of [22], which would result in the calculation of the stream order of Martian valley networks. With the determination of the stream order of Martian valley networks comes a greater understanding of the presence of surface liquids on the Martian surface, and therefore a greater understanding of the climate of Mars over its history [20], [22].

\subsubsection{Mapping polar-layered deposits and their unconformities}
Section 2.2.4 summarizes the currently unmapped polar-layered deposits (PLD) at the Martian polar caps. The PLDs captured by the Mariner 9 and Viking missions are unmapped, and although appearing to be random, may have cyclicity. Determining the cyclicity of the PLD could uncover previous Martian climate change [24], [25], as well as indicate patterns of Martian weather conditions such as dust storms [25], which would add to the total understanding of the Martian climate and atmosphere.

Also described in 2.2.4 are the unconformities present within the PLDs, created when erosion causes PLD layers to interrupt or overlay one another. Several of these unconformities have been captured during the Viking mission to span the PLDs on the polar caps of Mars [21], [44], but remain currently unmapped. Mapping the locations and patterning the disposition of all PLD unconformities would provide information about how the Martian climate and PLDs have changed over the planet’s history [21], [44].

However, manually mapping PLDs and their unconformities is slow, labor-intensive, and biased depending on the mapper [45].  Using machine learning patterning techniques, which are amenable to the autonomous mapping and patterning of PLDs and their unconformities, can counter the pitfalls of manual PLD mapping. Reference [45] proposes machine learning techniques such as clustering or classification to autonomously map planetary topography data, which have above 86\% accuracy in identifying and patterning terrain features. The cyclicities and patterns of PLD and their unconformities can be determined if these machine learning techniques are applied to image data of the dispositions of PLDs and their unconformities. In addition, automated PLD mapping eliminates the time and labor required for manual mapping. [46] also presents a neural network and clustering method of applying machine learning to the autonomous mapping of terrain pattern discovery that is above approximately 80\% accurate, but [46] uses the machine learning methods for the surface analysis and mapping of mineral deposits in the deep-sea floor. Similar to [45], the autonomous analysis of the surface terrain to determine patterning of [46] would save time and labor if utilized to pattern and map the disposition of PLDs and their unconformities.

\subsubsection{Analysis of MOMA data}
The Mars Organic Molecule Analyzer (MOMA) is a dual-source (laser desorption and gas chromatography) mass-spectrometer that will launch onboard the European Space Agency’s ExoMars rover in 2022 to analyze soil samples for signs of past or current life on the surface or subsurface of Mars [47], [48]. However, low bandwidth and high time of interplanetary data transfer will limit the downlink of all raw MOMA-collected data to Earth [48]. In addition, the soil samples are time-sensitive, so scientists will have a finite time to determine how to adjust ExoMars instruments to collect and study further data. To analyze collected mass-spectrometer data to prioritize the downlink of important or time-sensitive data, [48] has developed a machine learning approach to be trained using future MOMA data. The mass-spectrometry-focused neural network algorithm trained using MOMA modeled data will evaluate future MOMA data collected in-situ by the ExoMars rover, adjust rover instrument usage, and return the best or most time-sensitive data [50]. Harnessing machine learning for the analysis of MOMA data will allow only the most desirable data to be downlinked to Earth for analysis, and allow the ExoMars rover to make its own decisions on how to select and utilize its instruments to analyze soil sample data in its search for current or previous life on Mars [48].

\section{Conclusion and future scope}
This paper has outlined why machine learning methods would be beneficial for the exploration of Mars. Machine learning is already in use across numerous spacecraft purposed for Mars exploration to prioritize data selection, perform data collection, and analyze data. In addition, machine learning techniques are utilized on Earth computers to analyze raw Martian data. Further utilization of machine learning techniques in Mars spacecraft can expand the capabilities of Mars scientific missions by improving path-planning to save spacecraft energy, energy prediction to improve spacecraft path planning, which can potentially increase the capabilities of current and future Mars exploration missions. In addition, because automation would eliminate the need for manual data analysis, further applications of machine learning techniques to analyze Martian data would speed up the analysis process of Mars imagery or Mars-collected samples, reducing unproductivity and saving time. Most importantly, further applications of machine learning techniques for in-situ Mars exploration overcomes the issue of human hazards for the exploration of Mars.

\section{Acknowledgement}
I want to express my gratitude to Professor Bethany L. Ehlmann of the California Institute of Technology for providing numerous suggestions and resources about the features of Mars, and indicating the amenability of machine learning techniques to those features. She established a strong foundation for me to build upon, which I cannot be more grateful and appreciative of.

%----------------------------------------------------------------------------------------
%	REFERENCE LIST
%----------------------------------------------------------------------------------------

%----------------------------------------------------------------------------------------


\begin{thebibliography}{48} % Bibliography - this is intentionally simple in this template

{\footnotesize\bibliography{bibfile}}

%\bibitem[1]{Figueredo:2009dg}
%Figueredo, A.~J. and Wolf, P. S.~A. (2009).
%\newblock Assortative pairing and life history strategy - a cross-cultural study.
%\newblock {\em Human Nature}, 20:317--330.

\bibitem[1]{IN} I. El Naqa and M. J. Murphy, “What Is Machine Learning?,” Machine Learning in Radiation Oncology. Springer International Publishing, pp. 3–11, 2015. doi:10.1007/978-3-319-18305-3\_1.

\bibitem[2]{AG} A. McGovern and K. L. Wagstaff, “Machine learning in space: extending our reach,” Machine Learning, vol. 84, no. 3. Springer Science and Business Media LLC, pp. 335–340, Apr. 30, 2011. doi: 10.1007/s10994-011-5249-4.

\bibitem[3]{GG}G. Genta, “Reasons for human Mars exploration,” Next Stop Mars. Springer International Publishing, pp. 38–52, Dec. 31, 2016. doi: 10.1007/978-3-319-44311-9\_2.

\bibitem[4]{CL}C. Leovy, “Weather and climate on Mars,” Nature, vol. 412, no. 6843. Springer Science and Business Media LLC, pp. 245–249, Jul. 2001. doi: 10.1038/35084192.

\bibitem[5]{NB}N. Barlow, “Mars: An Introduction to its Interior, Surface and Atmosphere.” Cambridge University Press, 2008. doi: 10.1017/cbo9780511536069.

\bibitem[6]{DHM}D. M. Hassler et al., “Mars’ Surface Radiation Environment Measured with the Mars Science Laboratory’s Curiosity Rover,” Science, vol. 343, no. 6169. American Association for the Advancement of Science (AAAS), Jan. 24, 2014. doi: 10.1126/science.1244797.

\bibitem[7]{BMJ}B. M. Jakosky and R. J. Phillips, “Mars’ volatile and climate history,” Nature, vol. 412, no. 6843. Springer Science and Business Media LLC, pp. 237–244, Jul. 2001. doi: 10.1038/35084184.

\bibitem[8]{TE}T. Estlin et al., “Automated Targeting for the MER Rovers,” 2009 Third IEEE International Conference on Space Mission Challenges for Information Technology. IEEE, Jul. 2009. doi: 10.1109/smc-it.2009.38.

\bibitem[9]{MGT}M. G. Trainer et al., “Seasonal Variations in Atmospheric Composition as Measured in Gale Crater, Mars,” Journal of Geophysical Research: Planets, vol. 124, no. 11. American Geophysical Union (AGU), pp. 3000–3024, Nov. 2019. doi: 10.1029/2019je006175.

\bibitem[10]{LMC}L. M. Calle, W. Li, J. W. Buhrow, M. R. Johansen, and C. I. Calle, “Corrosion on Mars: An Investigation of Corrosion under Relevant Simulated Martian Environments,” in 48th International Conference on Environmental Systems, 2018. 

\bibitem[11]{DMH}D. M. Hassler et al., “Mars’ Surface Radiation Environment Measured with the Mars Science Laboratory’s Curiosity Rover,” Science, vol. 343, no. 6169. American Association for the Advancement of Science (AAAS), Jan. 24, 2014. doi: 10.1126/science.1244797.

\bibitem[12]{JAW}J. A. Whiteway et al., “Mars Water-Ice Clouds and Precipitation,” Science, vol. 325, no. 5936. American Association for the Advancement of Science (AAAS), pp. 68–70, Jul. 03, 2009. doi: 10.1126/science.1172344.

\bibitem[13]{HW}H. Wang, “Martian clouds observed by Mars Global Surveyor Mars Orbiter Camera,” Journal of Geophysical Research, vol. 107, no. E10. American Geophysical Union (AGU), 2002. doi: 10.1029/2001je001815.

\bibitem[14]{MAM}M. A. Mischna and S. Piqueux, “The role of atmospheric pressure on Mars surface properties and early Mars climate modeling,” Icarus, vol. 342. Elsevier BV, p. 113496, May 2020. doi: 10.1016/j.icarus.2019.113496.

\bibitem[15]{SC}S. Clifford, “The State and Future of Mars Polar Science and Exploration,” Icarus, vol. 144, no. 2. Elsevier BV, pp. 210–242, Apr. 2000. doi: 10.1006/icar.1999.6290.

\bibitem[16]{CRW}C. R. Webster et al., “Mars methane detection and variability at Gale crater,” Science, vol. 347, no. 6220. American Association for the Advancement of Science (AAAS), pp. 415–417, Dec. 16, 2014. doi: 10.1126/science.1261713.

\bibitem[17]{CO}C. Oze, “Have olivine, will gas: Serpentinization and the abiogenic production of methane on Mars,” Geophysical Research Letters, vol. 32, no. 10. American Geophysical Union (AGU), 2005. doi: 10.1029/2005gl022691

\bibitem[18]{SEL}S. E. Lauro et al., “Multiple subglacial water bodies below the south pole of Mars unveiled by new MARSIS data,” Nature Astronomy, vol. 5, no. 1. Springer Science and Business Media LLC, pp. 63–70, Sep. 28, 2020. doi: 10.1038/s41550-020-1200-6.

\bibitem[19]{Committee}Committee on Precursor Measurements Necessary to Support Human Operations on the Surface of Mars, “Physical Environmental Hazards” in Safe on Mars: Precursor Measurements Necessary to Support Human Operations on the Martian Surface, 2002.

\bibitem[20]{RZ} R. Zurek, “Polar Layered Terrains: Links Between the Martian Volatile and Dust Cycles,” in The 5th International Conference on Mars, 1999.

\bibitem[21]{SMM}S. M. Milkovich, “North polar cap of Mars: Polar layered deposit characterization and identification of a fundamental climate signal,” Journal of Geophysical Research, vol. 110, no. E1. American Geophysical Union (AGU), 2005. doi: 10.1029/2004je002349.

\bibitem[22]{AG}A. Gleyzer, M. Denisyuk, A. Rimmer, and Y. Salingar, “A FAST RECURSIVE GIS ALGORITHM FOR COMPUTING STRAHLER STREAM ORDER IN BRAIDED AND NONBRAIDED NETWORKS,” Journal of the American Water Resources Association, vol. 40, no. 4. Wiley, pp. 937–946, Aug. 2004. doi: 10.1111/j.1752-1688.2004.tb01057.x.

\bibitem[23]{FS}F. Stepinski, M. P. Mendenhall, and B. D. Bue, “Machine cataloging of impact craters on Mars,” Icarus, vol. 203, no. 1. Elsevier BV, pp. 77–87, Sep. 2009. doi: 10.1016/j.icarus.2009.04.026.

\bibitem[24]{BS}B. Smith, N. E. Putzig, J. W. Holt, and R. J. Phillips, “An ice age recorded in the polar deposits of Mars,” Science, vol. 352, no. 6289. American Association for the Advancement of Science (AAAS), pp. 1075–1078, May 27, 2016. doi: 10.1126/science.aad6968.

\bibitem[25]{KH}K. Herkenhoff, “Surface Ages and Resurfacing Rates of the Polar Layered Deposits on Mars,” Icarus, vol. 144, no. 2. Elsevier BV, pp. 243–253, Apr. 2000. doi: 10.1006/icar.1999.6287.

\bibitem[26]{RC}R. Castano et al., “Oasis: Onboard autonomous science investigation system for opportunistic rover science,” Journal of Field Robotics, vol. 24, no. 5. Wiley, pp. 379–397, 2007. doi: 10.1002/rob.20192.

\bibitem[27]{RF}R. Francis et al., “AEGIS autonomous targeting for ChemCam on Mars Science Laboratory: Deployment and results of initial science team use,” Science Robotics, vol. 2, no. 7. American Association for the Advancement of Science (AAAS), Jun. 28, 2017. doi: 10.1126/scirobotics.aan4582.

\bibitem[28]{NA}N. Abcouwer et al., “Machine Learning Based Path Planning for Improved Rover Navigation,” 2021 IEEE Aerospace Conference (50100). IEEE, Mar. 06, 2021. doi: 10.1109/aero50100.2021.9438337.

\bibitem[29]{KO}K. Otsu, G. Matheron, S. Ghosh, O. Toupet, and M. Ono, “Fast approximate clearance evaluation for rovers with articulated suspension systems,” Journal of Field Robotics, vol. 37, no. 5. Wiley, pp. 768–785, Jul. 09, 2019. doi: 10.1002/rob.21892.

\bibitem[30]{AC}A. Castano et al., “Automatic detection of dust devils and clouds on Mars,” Machine Vision and Applications, vol. 19, no. 5–6. Springer Science and Business Media LLC, pp. 467–482, Jun. 20, 2007. doi: 10.1007/s00138-007-0081-3.

\bibitem[31]{GD}G. Doran et al., “COSMIC: Content-based Onboard Summarization to Monitor Infrequent Change,” 2020 IEEE Aerospace Conference. IEEE, Mar. 2020. doi: 10.1109/aero47225.2020.9172337.

\bibitem[32]{MD}M. Dundar, B. L. Ehlmann, and E. Leask, “Machine-Learning-Driven New Geologic Discoveries at Mars Rover Landing Sites: Jezero Crater and NE Syrtis.” Wiley, Dec. 18, 2019. doi: 10.1002/essoar.10501294.1

\bibitem[33]{BR}B. Rothrock, R. Kennedy, C. Cunningham, J. Papon, M. Heverly, and M. Ono, “SPOC: Deep Learning-based Terrain Classification for Mars Rover Missions,” AIAA SPACE 2016. American Institute of Aeronautics and Astronautics, Sep. 09, 2016. doi: 10.2514/6.2016-5539.

\bibitem[34]{MO}M. Ono et al., “MAARS: Machine learning-based Analytics for Automated Rover Systems,” 2020 IEEE Aerospace Conference. IEEE, Mar. 2020. doi: 10.1109/aero47225.2020.9172271

\bibitem[35]{RMS}R. M. Swan et al., “AI4MARS: A Dataset for Terrain-Aware Autonomous Driving on Mars,” 2021 IEEE/CVF Conference on Computer Vision and Pattern Recognition Workshops (CVPRW). IEEE, Jun. 2021. doi: 10.1109/cvprw53098.2021.00226.

\bibitem[36]{RB}R. Boumghar, L. Lucas, and A. Donati, “Machine Learning in Operations for the Mars Express Orbiter,” 2018 SpaceOps Conference. American Institute of Aeronautics and Astronautics, May 25, 2018. doi: 10.2514/6.2018-2551.

\bibitem[37]{MP}M. Petkovic et al., “Machine Learning for Predicting Thermal Power Consumption of the Mars Express Spacecraft,” IEEE Aerospace and Electronic Systems Magazine, vol. 34, no. 7. Institute of Electrical and Electronics Engineers (IEEE), pp. 46–60, Jul. 01, 2019. doi: 10.1109/maes.2019.2915456.

\bibitem[38]{DQ}D. Qiu et al., “SCOTI: Science Captioning of Terrain Images for data prioritization and local image search,” Planetary and Space Science, vol. 188. Elsevier BV, p. 104943, Sep. 2020. doi: 10.1016/j.pss.2020.104943.

\bibitem[39]{SH}S. Higa et al., “Vision-Based Estimation of Driving Energy for Planetary Rovers Using Deep Learning and Terramechanics,” IEEE Robotics and Automation Letters, vol. 4, no. 4. Institute of Electrical and Electronics Engineers (IEEE), pp. 3876–3883, Oct. 2019. doi: 10.1109/lra.2019.2928765.

\bibitem[40]{SG}S. Ghosh, K. Otsu, and M. Ono, “Probabilistic Kinematic State Estimation for Motion Planning of Planetary Rovers,” 2018 IEEE/RSJ International Conference on Intelligent Robots and Systems (IROS). IEEE, Oct. 2018. doi: 10.1109/iros.2018.8593771.

\bibitem[41]{UM} U. Michelucci, M. Baumgartner, and F. Venturini, “Optical Oxygen Sensing with Artificial Intelligence,” Sensors, vol. 19, no. 4. MDPI AG, p. 777, Feb. 14, 2019. doi: 10.3390/s19040777.

\bibitem[42]{WJ}W. Ji et al., “Extreme Learning Machine for Robustness Enhancement of Gas Detection Based on Tunable Diode Laser Absorption Spectroscopy.” MDPI AG, Dec. 28, 2018. doi: 10.20944/preprints201812.0331.v1.

\bibitem[43]{AR}A. Rashno, B. Nazari, S. Sadri, and M. Saraee, “Effective pixel classification of Mars images based on ant colony optimization feature selection and extreme learning machine,” Neurocomputing, vol. 226. Elsevier BV, pp. 66–79, Feb. 2017. doi: 10.1016/j.neucom.2016.11.030.

\bibitem[44]{IMTF} I. Molloy and T. F. Stepinski, “Automatic mapping of valley networks on Mars,” Computers \& Geosciences, vol. 33, no. 6. Elsevier BV, pp. 728–738, Jun. 2007. doi: 10.1016/j.cageo.2006.09.009.

\bibitem[45]{TF}T. F. and R. Vilalt, “Machine Learning Tools for Geomorphic Mapping of Planetary Surfaces,” Machine Learning. InTech, Feb. 01, 2010. doi: 10.5772/9146.

\bibitem[46]{CJ}C. Juliani and E. Juliani, “Deep learning of terrain morphology and pattern discovery via network-based representational similarity analysis for deep-sea mineral exploration,” Ore Geology Reviews, vol. 129. Elsevier BV, p. 103936, Feb. 2021. doi: 10.1016/j.oregeorev.2020.103936.

\bibitem[47]{FG}F. Goesmann et al., “The Mars Organic Molecule Analyzer (MOMA) Instrument: Characterization of Organic Material in Martian Sediments,” Astrobiology, vol. 17, no. 6–7. Mary Ann Liebert Inc, pp. 655–685, Jul. 2017. doi: 10.1089/ast.2016.1551.

\bibitem[48]{VDP}V. Da-Poian, E. Lyness, W. Brinckerhoff, R. Danell, X. Li, and M. Trainer, “Science Autonomy and the ExoMars Mission: Machine Learning to Help Find Life on Mars,” Goldschmidt Abstracts. Geochemical Society, 2020. doi: 10.46427/gold2020.522.

\bibitem[49]{IP}I. Priyadarshini and V. Puri, “Mars weather data analysis using machine learning techniques,” Earth Science Informatics, vol. 14, no. 4. Springer Science and Business Media LLC, pp. 1885–1898, Jul. 04, 2021. doi: 10.1007/s12145-021-00643-0.

 
\end{thebibliography}
\end{document}